\documentclass[%
 reprint,
superscriptaddress,
showpacs,
 amsmath,amssymb,?
 aps,
 pra,
]{revtex4-1}

\usepackage{graphicx}
\usepackage{dcolumn}
\usepackage{bm}
\usepackage[dvips]{color}

\usepackage{float}

\newcommand{\sech}{\mathrm{sech}}

\usepackage{color}

\def\tOmega{\tilde\Omega}
\def\tD{\tilde \Delta}   \def\tLs{\tilde{\Lambda}_s}

\begin{document}

\title{Adiabatic tracking for photo- and magneto-association of  Bose-Einstein condensates with Kerr nonlinearities}
\author{M. Gevorgyan}
\affiliation{Laboratoire Interdisciplinaire Carnot de Bourgogne, CNRS UMR 6303, Universit\'e de Bourgogne Franche Comt\'e,
BP 47870, 21078 Dijon, France}
\affiliation{Institute for Physical Research NAS of Armenia - 0203 Ashtarak-2, Armenia}
\author{S. Gu\'erin}
\email{sguerin@u-bourgogne.fr}
\affiliation{Laboratoire Interdisciplinaire Carnot de Bourgogne, CNRS UMR 6303, Universit\'e de Bourgogne Franche Comt\'e,
BP 47870, 21078 Dijon, France}
\author{C. Leroy}
\affiliation{Laboratoire Interdisciplinaire Carnot de Bourgogne, CNRS UMR 6303, Universit\'e de Bourgogne Franche Comt\'e,
BP 47870, 21078 Dijon, France}
\author{A. Ishkhanyan}
\affiliation{Institute for Physical Research NAS of Armenia - 0203 Ashtarak-2, Armenia}
\affiliation{Armenian State Pedagogical University, Yerevan 0010, Armenia}
\affiliation{Institute of Physics and Technology, National Research Tomsk Polytechnic University, Tomsk 634050, Russia}
\author{H. R. Jauslin}
\affiliation{Laboratoire Interdisciplinaire Carnot de Bourgogne, CNRS UMR 6303, Universit\'e de Bourgogne Franche Comt\'e,
BP 47870, 21078 Dijon, France}

\begin{abstract}
We develop the method of adiabatic tracking for photo- and magneto-association of Bose-Einstein atomic condensates with models that include Kerr type nonlinearities. We show that the inclusion of these terms can produce qualitatively important modifications in the adiabatic dynamics, like the appearance of bifurcations, in which the trajectory that is being tracked loses its stability. As a consequence
the adiabatic theorem does not apply and the adiabatic transfer can be strongly degraded. This degradation
can be compensated by using fields that are strong enough compared with the values of the Kerr terms. The main result is that, despite these potentially detrimental features, there is always a choice of the detuning that leads to an efficient adiabatic tracking, even for relatively weak fields.
\end{abstract}
\maketitle

\section{Introduction}

The formation of molecules from ultracold atom gases by external fields and more generally the coherent oscillations between the atomic and molecular Bose-Einstein condensates (BEC) can be well described in the mean-field approximation by an effective quantum system featuring nonlinearites \cite{Pitaevskii,Javan,Drummond,Timmermans,Kohler,CChin}.  More specifically a driven non-linear two-level model is already a very good approximation accounting for the one-color photo-association or for magneto-association via a by Feshbach resonance.

The popular adiabatic passage techniques in quantum mechanics \cite{adiab} need a reformulation for nonlinear quantum systems: The Schr\"odinger equation is formally reinterpreted as classical Hamilton equations of motion for which the classical adiabatic theorem can be applied in the underlying phase space.
The adiabatic trajectory is formed by the instantaneous (elliptic) fixed points defined at each value of the adiabatic parameters and continuously connected to the initial condition. Obstructions of classical adiabatic passage are given by
the presence of a separatrix in the vicinity of the instantaneous fixed point which involves arbitrary small frequencies \cite{Arnold-Kozlov-Neihstadt-book,Henrard-adiabatic-book,Cary-Escande-Tennyson-adiabatic-1986}.
The adiabatic evolution ultimately breaks down when there is a bifurcation, in which the followed stable elliptic fixed point crosses an unstable hyperbolic fixed point and becomes itself hyperbolic.

Recently inverse engineering methods have been introduced, which allow one to derive the parameters of the driving field by forcing the system to reach a given target via shortcut to adiabaticity \cite{shortcut} and by imposing additionally robustness features \cite{robust1,robust2}.

In a recent article \cite{Guerin-etal-tracking-1} a method of adiabatic tracking was proposed for the photo- and magneto-association of atomic  BEC, in which for a given pulse shape of the driving field and a chosen time evolution of the population, a profile of the detuning (defined as the difference of the frequency of the two-atom-molecule transition and the chirped laser frequency) can be constructed that  provides a close tracking of the population in the adiabatic regime. The main point in this approach is  that the tracking strategy prevents the crossing of a separatrix that was the main source of uncontrolled population evolution leading to the decreasing of the final conversion to the molecular state. The method was analyzed for a simplified model, which did not contain the Kerr terms stemming from elastic interparticles scattering. The goal of the present work is to extend the analysis to include the Kerr terms [see Eqs. \eqref{model}].

The main result is that in the presence of Kerr terms it is still possible to construct a detuning  that leads to an efficient tracking of the chosen population dynamics despite the strong modifications of the structure of the adiabatic phase portrait. The main feature is the crossing of fixed point trajectories, which induces bifurcations and loss of stability of the adiabatic dynamics.

After having introduced the model in Sect. II, we show in Sect. III that the presence of Kerr terms has some strong qualitative effects on the dynamics, like the appearance of other hyperbolic points that can interfere with the desired tracking.  A detailed analysis of these features leads to the determination of a detuning that produces efficient tracking. We conclude in Sect. IV.

\section{The model}

The model including second-order nonlinearities and third-order Kerr nonlinearities is defined by the equations
\cite{Pitaevskii,Javan,Drummond,Timmermans,Kohler,Garraway,Ishkhanyan}
\begin{subequations}
\label{model}
\begin{eqnarray}
\label{eq: ia1}
i\dot a_{1} &=&\frac{ \Omega}{\sqrt{2}}e^{-i\int^{t}_{t_i}\Delta(t')dt'}\bar a_{1}a_{2}+(\Lambda_{11}|a_{1}|^{2}+\Lambda_{12}|a_{2}|^{2})a_{1}\qquad\\
\label{eq: ia2}
i\dot a_{2} &=&\frac{\Omega}{2\sqrt{2}}e^{i\int^{t}_{t_i}\Delta(t')dt'}a_{1}^2 +(\Lambda_{21}|a_{1}|^{2}+\Lambda_{22}|a_{2}|^{2})a_{2},
\end{eqnarray}
\end{subequations}
where $a_{1}$ and $\sqrt{2} a_{2}$ are the atomic and molecular state probability amplitudes, respectively, with $|a_{1}|^{2}+2|a_{2}|^{2}=1$, ${ \Omega}(t)$ is the Rabi frequency, which is chosen to be  real and positive $\Omega\geq 0$, and $\Delta(t)$ is the detuning. The bar denotes the complex conjugate.
The third-order nonlinearities  $\Lambda_{11},$ $\Lambda_{12}=\Lambda_{21}$, and $\Lambda_{22}$ describe atom-atom,
atom-molecule and molecule-molecule elastic scatterings, respectively.
The initial condition, at $t_i=-\infty$, is  the all-atomic state:
$$|a_{1}(-\infty)|=1, \qquad |a_{2}(-\infty)|=0.
$$
Using the transformation $a_{1}(t)=c_{1}(t)e^{-i\int^t  ds{\Delta(s)}/{3}}$, $a_{2}(t)=c_{2}(t)e^{i\int^t ds {\Delta(s)}/{3}}$,
 we can write
\begin{subequations}
 \label{eq: ic1-2}
 \begin{eqnarray}
 \label{eq: ic1}
i\dot c_{1} & = &\left[-\frac{\Delta}{3}+\Lambda_{11}|c_{1}|^{2}+\Lambda_{12}|c_{2}|^{2}\right]c_{1}+\frac{\Omega}{\sqrt{2}}\bar c_{1}c_{2} \\ \label{eq: ic2}
i\dot c_{2}  & = & \left[\frac{\Delta}{3}+\Lambda_{21}|c_{1}|^{2}+\Lambda_{22}|c_{2}|^{2}\right]c_{2}+\frac{\Omega}{2\sqrt{2}}c_{1}^2.
\end{eqnarray}
\end{subequations}

\subsection{Analysis of the model with Kerr terms}
One can rewrite the preceding equations in order to reveal two relevant combinations of Kerr terms. The (population preserving) transformation
\begin{subequations}
 \begin{eqnarray}
b_{1}(t)&=&c_{1}(t)e^{i\int^t  ds[{\Lambda_{11}\vert c_1(s)\vert^2+\Lambda_{12}\vert c_2(s)\vert^2-\Delta(s)}/{3}]},\\
b_{2}(t)&=&c_{2}(t)e^{2i\int^t  ds[{\Lambda_{11}\vert c_1(s)\vert^2+\Lambda_{12}\vert c_2(s)\vert^2-\Delta(s)}/{3}]}
\end{eqnarray}
\end{subequations}
leads to
\begin{subequations}
 \label{eq: ib1-2}
 \begin{eqnarray}
i\dot b_{1} & = &\frac{\Omega}{\sqrt{2}}\bar b_{1}b_{2} \\ \label{eq: ib2}
i\dot b_{2}  & = & \left[\Delta-\Lambda_a+2\Lambda_{s}|b_{2}|^{2}\right]b_{2}+\frac{\Omega}{2\sqrt{2}}b_{1}^2
\end{eqnarray}
\end{subequations}
with the notations
\begin{equation}\label{eq: 222}
\Lambda_{s}=2\Lambda_{11}+\frac{\Lambda_{22}}{2}-2\Lambda_{12}, \qquad \Lambda_{a}=2\Lambda_{11}-\Lambda_{12}.
\end{equation}
Considering the example of a $^{87}$Rb condensate, we have $\Lambda_s>0$ \cite{Drummond2,Mackie-2005}.
We remark that these equations depend only on the combinations $\Lambda_a$ and $\Lambda_s$. The Kerr terms produce two distinct effects: The term $\Lambda_a$ produces only a constant shift in the detuning, which can be trivially compensated in the adiabatic tracking,
while the term $\Lambda_s$ produces a non-linear term in the coupling. Therefore, only one effective parameter $\Lambda_s$ has to be taken into account in the analysis.

We explore in this paper regimes for a Rabi frequency up to the same order as the Kerr terms, which will allow the use of weak fields. We show that a specifically shaped detuning can lead to high-efficient transfer.

The adiabatic treatment of the nonlinear model, which allows the extension of the standard adiabatic passage of linear models, can be accomplished by a classical Hamiltonian formulation of the problem. The equations of motion \eqref{eq: ic1-2} can be written in Hamiltonian form:
$$
i\frac{d c_{1}}{dt}=\frac{\partial h}{\partial \bar c_{1}}
,\qquad
i\frac{d c_{2}}{dt}=\frac{\partial h}{\partial \bar c_{2}},
$$
with  the Hamilton function
\begin{align}
\label{eq: h1}
&h=\frac{\Delta}{3}(|c_{2}|^{2}-|c_{1}|^{2})+\frac{\Omega}{2\sqrt{2}}(c_{1}^{2}\bar{c}_{2}+\bar{c}_{1}^{2}c_{2})\nonumber\\
&\qquad+\frac{\Lambda_{11}}{2}|c_{1}|^{4}+\frac{\Lambda_{22}}{2}|c_{2}|^{4}+\Lambda_{12}|c_{1}|^{2}|c_{2}|^{2}.
\end{align}
These equations are a complex coordinate formulation of a classical Hamiltonian system, where the standard real canonical coordinates $p_{j},$ $q_{j}$ are defined as the imaginary and real parts of the complex coefficients:
$
c_{j}=({q_{j}+ip_{j}})/{\sqrt{2}}.
$
For constant values of $\Omega$ and $\Delta$ this Hamiltonian system with two degrees of freedom is Liouville integrable, since $J:=  |c_1|^2+2|c_2|^2$ is a conserved quantity,
which allows one to reduce the dimension of the phase space by 2, leading to an effective one degree of freedom system.
In fact this system can be viewed as an example of a classical  $1:2$ Fermi resonance Hamiltonian normal form \cite{Efstathiou-book,Cushman-Bates-book}. This reduction is singular: the reduced phase space is like a sphere with a conical singularity as depicted in Figs 1-4. This fact plays an important role in the dynamics of the system.

\par
In order to analyze its properties we first perform a canonical  variable transformation
$(p_j,q_j)\mapsto (I_{j},\varphi_{j})$  defined by $c_{j}=\sqrt{I_{j}}e^{-i\varphi_{j}}$, and
further, a second canonical transformation $(I_{1}, \varphi_{1}, I_{2}, \varphi_{2})\mapsto(I, \alpha, J, \gamma)$ defined by
\begin{subequations}
\begin{eqnarray}
\gamma&=&\varphi_{1}, \qquad J=I_{1}+2I_{2}\\
\alpha&=&-2\varphi_{1}+\varphi_{2}, \qquad I=I_{2}
\end{eqnarray}
\end{subequations}
In these variables, the Hamilton function  \eqref{eq: h1} reads
\begin{equation}\label{eq: h2new}
h=(\Delta-\Lambda_aJ) I+\Lambda_s I^{2}+\frac{\Omega}{\sqrt{2}}(J-2I)\sqrt{I}\cos{\alpha}-C
\end{equation}
with $C=\frac{\Delta}{3}J-\frac{\Lambda_{11}}{2}J^2$. The term $\Lambda_s$ adds a quadratic term in the Hamilton function.

Since the Hamilton function is independent of $\gamma$, we can define a reduced phase space of only two
dimensions with the variables $(I, \alpha)$, satisfying $\dot\alpha=\partial h/\partial I$ and $\dot I=-\partial h/\partial \alpha$.
 Instead of the canonical variable $I$ it is convenient to use the variable $P:=2I=2|c_2|^2$, which has the physical interpretation of the probability for the system to be in the molecular state.
The equations of motion for these variables, with $J=1$, are
\begin{subequations}
\label{eq: I2new-alfa2new}
\begin{eqnarray}\label{eq: I2new}
\frac{dP}{dt}& = &\Omega(1-P)\sqrt{P}\sin \alpha\\ \label{eq: alfa2new}
\frac{d\alpha}{dt}& = &(\Delta-\Lambda_a)+\Lambda_s P+\Omega\frac{(1-3 P)}{2\sqrt{P}} \cos\alpha.
\end{eqnarray}
\end{subequations}
We notice that the Kerr terms appear only in the second equation \eqref{eq: alfa2new}. As a consequence, from the first equation we can deduce the following result, that was stated for the model without Kerr terms \cite{Guerin-etal-tracking-1}:
 For the initial condition $P(t_{i})=0$, from Eq. \eqref{eq: I2new} we obtain
\begin{equation}\label{eq: p}
P(t)=\tanh^{2}\left[\int_{t_{i}}^{t}\frac{\Omega(t')}{2}\sin \alpha(t')dt'\right].
\end{equation}
This formula shows that one cannot achieve a complete transfer with pulses of a finite area,  as it was already established for the case without Kerr terms.

We remark that the angle $\alpha$ is not well-defined for $I=0$ nor for $I=1/2$. In order to display a faithful picture of the reduced phase space one can use a different set of variables, as described e.g. in \cite{Efstathiou-book,Cushman-Bates-book}, which generalizes the coordinates (inversion and coherences) for the Bloch sphere to the non-linear problem:
\begin{eqnarray*}
J\equiv \Pi_0 :=&   |c_1|^2+2|c_2|^2 &=  1\\
\Pi_1 :=&   |c_1|^2-2|c_2|^2 &=   J-2P\\
\Pi_2 :=&  \null~~2 (c_1^{2}\bar c_2 + \bar c_1^{2}c_2 ) &=   2\sqrt{2}(J-P)\sqrt{P}\cos \alpha \\
\Pi_3 :=& -2i (c_1^{2}\bar c_2 - \bar c_1^{2}c_2 ) &=   2\sqrt{2}(J-P)\sqrt{P}\sin \alpha.
\end{eqnarray*}
These variables  satisfy the relation
\begin{equation}
\label{ }
\Pi_2^2+ \Pi_3^2 = 8(1-P)^2 P, \qquad P\in[0,1].
\end{equation}
This last equation represents the reduced phase space as a two-dimensional surface, embedded in a 3-dimensional space of coordinates $P,\Pi_2,\Pi_3$, that has the shape of a drop with a point singularity at  the top $(P=1, \Pi_2=0, \Pi_3=0)$,
depicted  in Figs. \ref{fig:Alfa0_0_5_Fixed_Points} to \ref{fig:AlfaPi_1_1_Fixed_Points}.
The Hamiltonian \eqref{eq: h1}  can be expressed as
$$
h= \frac{\Omega}{4\sqrt{2}} \Pi_2 + \frac{1}{2}(\Delta-\Lambda_a) P+ \frac{1}{4}\Lambda_s P^2-C
$$
and the equations of motion for these variables of the reduced phase space are given by
\begin{eqnarray*}
\frac{dP}{dt} &=&  \{P,h \} =   \frac{\Omega}{2\sqrt{2}}  \Pi_3 \\
\frac{d\Pi_2}{dt}  &=& \{\Pi_2,h \} =  - \Pi_3 (\Delta-\Lambda_a+ \Lambda_s P)\\
\frac{d\Pi_3}{dt} &=&  \{\Pi_3,h \} \\
&=&\sqrt{2} \Omega (1-P)(1-3P) + \Pi_2(\Delta-\Lambda_a+ \Lambda_s P),
\end{eqnarray*}
where the Poisson bracket  in the complex representation is given  by
 \def\pp(#1,#2){\frac{\partial #1}{\partial #2}}
\begin{eqnarray*}
  \{g, h\}& = & -\sum_{n} \pp(g,p_n)   \pp(h,q_n)-\pp(h,p_n)\pp(g,q_n) \\
&=&   -i \sum_{n} \pp(g,c_n)   \pp(h,\bar c_n)-\pp(h,c_n)\pp(g,\bar c_n).
\end{eqnarray*}

  \subsection{Reduced dimensionless variables}

  We will consider pulses of the form
  $$
  \Omega(t)=\Omega_0 	\tOmega(t/T)>0,
  $$
where $\tOmega({t}/{T})$ is a pulse shape function (e.g. $\tOmega(t/T)=\sech(t/T)$).
 Defining the following dimensionless variables
 \begin{equation}\label{dimensionless}
 s:= \Omega_0 t, \ \ \tLs:= \frac{\Lambda_s}{\Omega_0},\ \ \tD:=\frac{\Delta-\Lambda_a}{\Omega_0},
\  \ \tau:=\Omega_0 T,
\end{equation}
 the time evolution equations \eqref{eq: I2new-alfa2new} become
\begin{subequations}
\label{eq: I2alpha2reduced}
\begin{eqnarray}\label{eq: I2reduced}
\frac{dP}{ds}& = &\tOmega\left(\frac{s}{\tau}\right)(1-P)\sqrt{P}\sin \alpha, \\ \label{eq: alfa2reduced}
\frac{d\alpha}{ds}& = &\tD\left(\frac{s}{\tau}\right)+\tLs P+\tOmega\left(\frac{s}{\tau}\right)\frac{1-3 P}{2\sqrt{P}} \cos\alpha.
\end{eqnarray}
\end{subequations}
The dynamics is thus determined by two dimensionless parameters, $\tLs$ that describes the effect of the Kerr terms and $\tau$ that determines the adiabatic scale.

\section{Adiabatic tracking}

The classical adiabatic theorem for a system with one degree of freedom is generally presented  \cite{Arnold-Kozlov-Neihstadt-book,Henrard-adiabatic-book,Cary-Escande-Tennyson-adiabatic-1986} as the statement that the action variable is an adiabatic invariant, i.e. it stays constant in the asymptotic limit when the parameters of the system are slowly varying. The adiabatic theorem that we apply in this paper for Eqs.  \eqref{eq: I2alpha2reduced}, in the adiabatic limit $\tau\to \infty$, is a particular case of this general adiabatic theorem \cite{Fasano-Marmi-book}: if the initial condition is a stable fixed point, the value of the corresponding action variable is equal to zero. The statement that the value of the action variable stays constant implies in this case that the adiabatic evolution follows the instantaneous stable fixed points. The adiabatic evolution breaks down when there is a bifurcation, in which the followed stable elliptic fixed point crosses an unstable hyperbolic fixed point and becomes itself hyperbolic.


The idea of adiabatic tracking is to choose  a pulse shape
$\Omega(t)$ and a molecular  population evolution  $P_{\text{track}}(t)$ that one wishes to follow, and then to determine a detuning $\Delta_{\text{track}}(t)$ such that the exact solution corresponding to
 $\Omega(t), \Delta_{\text{track}}(t)$ approaches the desired  $P_{\text{track}}(t)$ in the adiabatic limit. In order to achieve this, first we determine $\Delta_{\text{track}}(t)$ such that an instantaneous fixed point of Eqs.  \eqref{eq: I2alpha2reduced},
  satisfies $P(t)=P_{\text{track}}(t)$.
As in \cite{Guerin-etal-tracking-1}, we will use for illustration pulse shapes of the form
 \begin{equation}\label{pulse-shape}
 \Omega(t)=\Omega_0\sech(t/T)
\end{equation}
and $P_{\text{track}}(t)$ of the form
 \begin{align}\label{sin2ex}
 P_{\text{track}}(t)&=\sin^2
\frac{1}{2T}\int_{-\infty}^t\sech(t'/T )dt'\nonumber\\
& = \sin^2 \bigl[ \arctan(\sinh(t/T))/2+\pi/4 \bigr].
\end{align}
 The second step is to verify if the dynamics defined by this $\Omega(t), \Delta_{\text{track}}(t)$ satisfies the conditions of validity of the adiabatic theorem. The main point is to verify whether there is a crossing of  the tracked fixed point (which starts out as a stable elliptic fixed point) and  other fixed points of hyperbolic type. In references
\cite{Liu-PRL2003,Itin-Watanabe-PRE2007,Itin-Vasiliev-Krishna-Watanabe,Itin-Torma-2009}
it was shown that a crossing of a separatrix associated to a hyperbolic fixed point can produce a large deviation from the adiabatic approximation, and a strongly random behavior (see also \cite{Itin-Watanabe-PRL2007,Itin-Watanabe-Konotop-2008}  for further developments in the subject). General methods to analyse the crossing of separatrices are described in \cite{Cary-Escande-Tennyson-adiabatic-1986,Arnold-Kozlov-Neihstadt-book,Henrard-adiabatic-book}.
In \cite{Guerin-etal-tracking-1} it was shown that in the model without Kerr terms the tracking solution can be chosen such that there are no crossings with other fixed points at finite times when the laser amplitude is non-zero, and thus the adiabatic approximation is justified and one obtains a good tracking with the desired behavior.

 \subsection{Fixed points and adiabatic  tracking formula}
In order to analyze the properties of the adiabatic tracking when the Kerr terms are included, we have to determine the instantaneous fixed points and the position of the instantaneous separatrices.

The instantaneous fixed points in the reduced phase space, for given $\Delta$ and $\Omega$, are obtained by setting  $\dot{P}=0, \dot{\alpha}=0$ in \eqref{eq: I2alpha2reduced}. The first equation \eqref{eq: I2reduced} has  two solutions: $\alpha=0$ and $\alpha=\pi$. Inserting  into \eqref{eq: alfa2reduced} we obtain an algebraic equation for the corresponding fixed point  populations $P_0$
and $P_{\pi}$:
\begin{equation}\label{Delta-last2}
\tD=-\tLs P_0-\frac{\tOmega}{2\sqrt{P_{0}}}(1-3P_{0}) \qquad\text{for} ~\alpha=0,
\end{equation}
and
\begin{equation}\label{Delta-last2pi}
\tD=-\tLs P_{\pi}+\frac{\tOmega}{2\sqrt{P_{\pi}}}(1-3P_{\pi}) \qquad \text{for}~ \alpha=\pi.
\end{equation}
One can try to use either of these equations  to construct a detuning, denoted $\tD_{\text{track}, 0}(s/\tau)$ and $\tD_{\text{track}, \pi}(s/\tau)$ respectively, for the desired tracking of a given $P_{\text{track}}(s/\tau)$:
\begin{equation}\label{trackingformula}
\tD_{\text{track}, 0}=-\tLs P_{\text{track}}-\frac{\tOmega}{2\sqrt{P_{\text{track}}}}(1-3P_{\text{track}})
\end{equation}
and
\begin{equation}\label{trackingformula-pi}
\tD_{\text{track}, \pi}=-\tLs P_{\text{track}}+\frac{\tOmega}{2\sqrt{P_{\text{track}}}}(1-3P_{\text{track}}).
\end{equation}
We remark that the points  $P=1$ and for $\Omega=0$, $P=0$  are also fixed points,  but since the coordinate $\alpha$ is not defined there, it has to be verified in the original coordinates. $P=0$ is an elliptic fixed point, and it corresponds to the initial condition, in which the condensate is in the all-atomic state. $P=1$ corresponds to the target all-molecular state. Its stability character can change during the time evolution from elliptic to hyperbolic, as we will see in the next section.

 \begin{figure*}[t] 
\begin{center}
\includegraphics[scale=0.95]{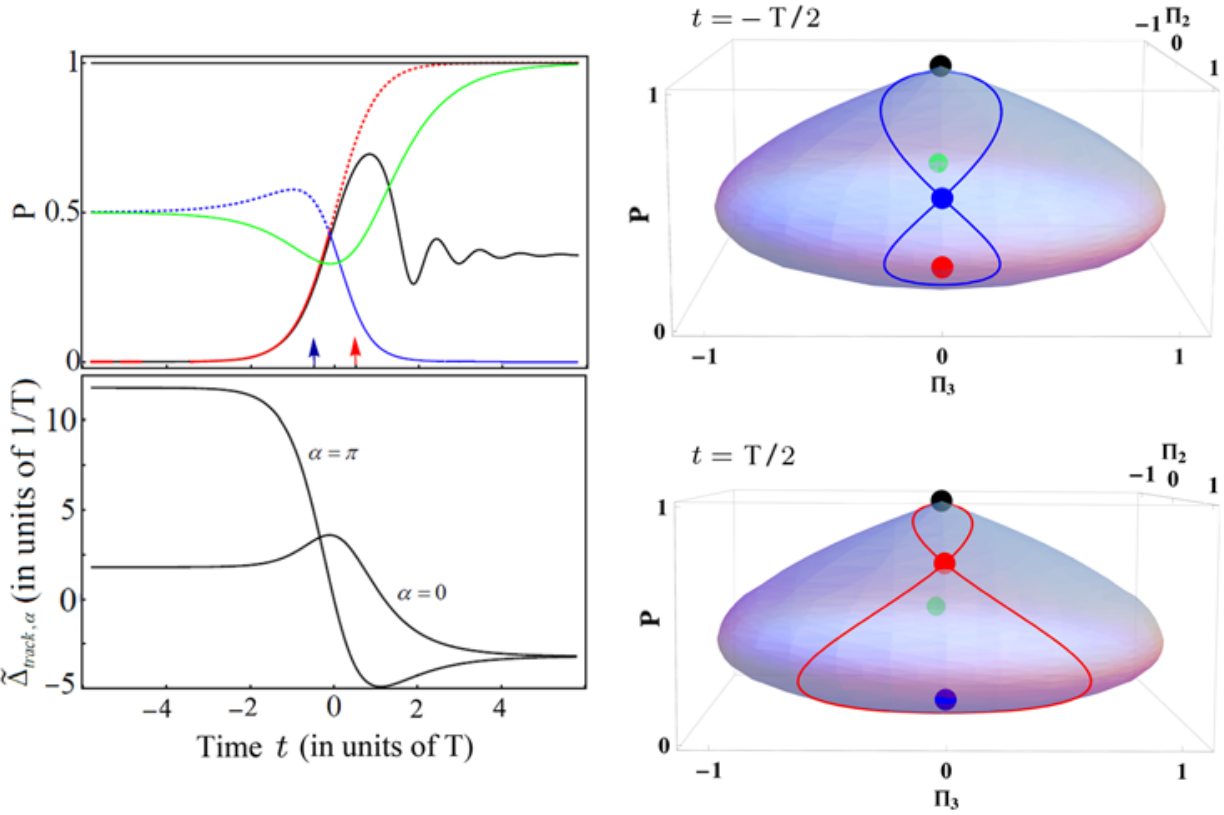}
\end{center}
\caption{Tracking with $\tD_{\text{track}, \alpha=0}$, $\Omega_0/\Lambda_s=0.5$, $\tau=5$.
Upper left frame (a): Exact (numerical) solution (black line), $P_{\text{track}}$ (red line), and fixed points $P_0$ (blue and green lines; elliptic: solid lines, hyperbolic: dashed lines) solutions of Eq. \eqref{fixed-points-0} (other than $P_{\text{track}}$), for the corresponding detuning [line $\alpha=0$, Eq. \eqref{trackingformula} shown in the lower left panel; the other line in this panel  corresponds to the other choice of detuning with $\alpha=\pi$, Eq. \eqref{trackingformula-pi}]. Right frame: Instantaneous phase portraits, with fixed points and separatrices,  at  the times indicated by the arrows in the left frame. The red and the blue curves are on the same side of phase space, and they intersect, while the green one is on the opposite side. The tracking fixed point (red dot) is elliptic for $t=-T/2$ [panel (b)] and hyperbolic for $t=T/2$ [panel (c)] after the bifurcation by crossing with the blue fixed point.}
\label{fig:Alfa0_0_5_Fixed_Points}
\begin{center}
\includegraphics[scale=0.95]{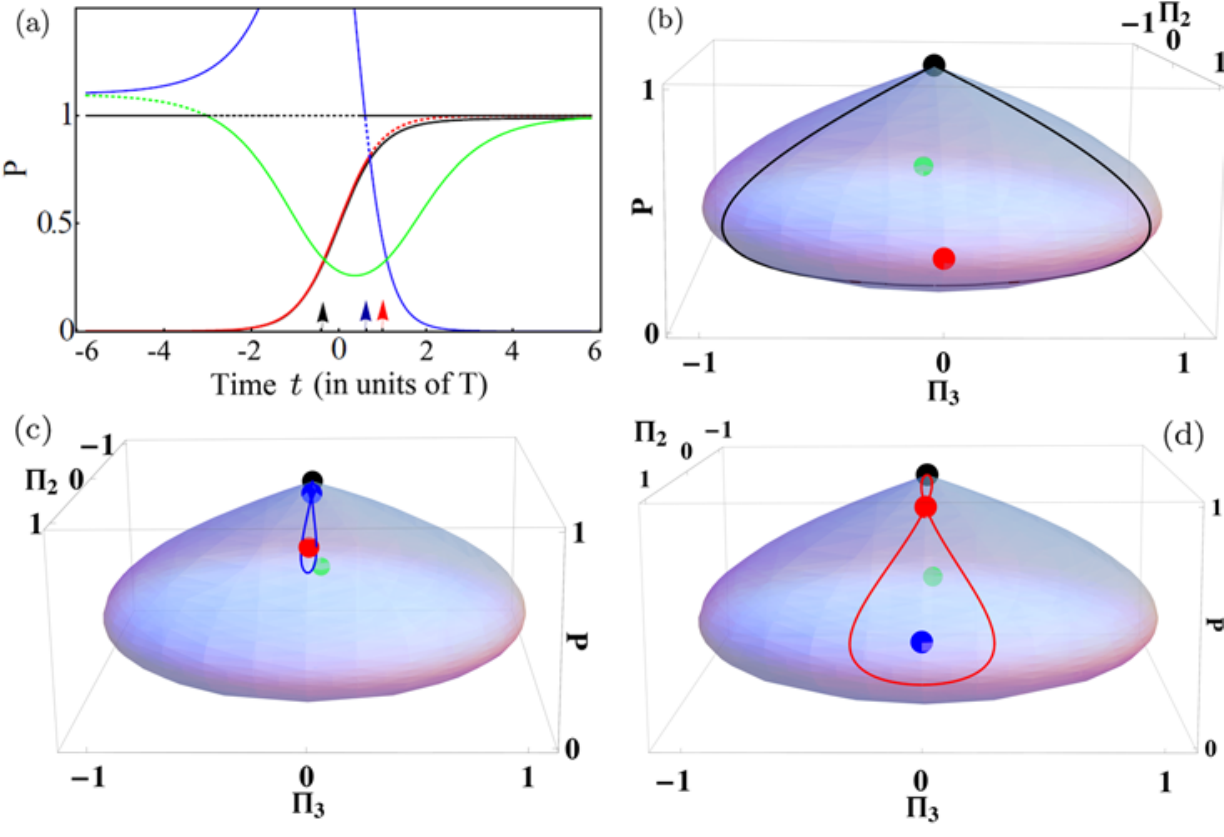}
\end{center}
\caption{Same as Fig. 1 but for $\Omega_0/\Lambda_s=1.1$ and $\tau=5.5$.  Panels (b), (c), and (d) display instantaneous phase portraits at  the times indicated by the arrows from left to right in frame (a), respectively.
The tracking fixed point (red dot) is elliptic in panels (b) and (c), and hyperbolic in (d), after the bifurcation by crossing with the blue fixed point.}
 \label{fig:Alfa0_1_1_Fixed_Points}
\end{figure*}

In the model without Kerr terms discussed in \cite{Guerin-etal-tracking-1} the two  choices \eqref{trackingformula} and
\eqref{trackingformula-pi} are essentially equivalent and they lead to the same quality of transfer. As we will show below, when Kerr terms are present the two choices lead to qualitatively different dynamics.
In the lower left panel of Fig. \ref{fig:Alfa0_0_5_Fixed_Points} we display an example of the two choices $\tD_{\text{track}, 0}$ and $\tD_{\text{track}, \pi}$.
The  second choice
\eqref{trackingformula-pi} leads to  significantly better transfer properties.  We will show indeed that for
\eqref{trackingformula} the tracking fixed point, which starts being elliptic, goes inevitably through an intersection with a hyperbolic fixed point, and its stability is lost by becoming in turn hyperbolic. Once the tracking fixed point is hyperbolic the classical adiabatic theorem does not apply anymore, and thus the final stages of the process are non-adiabatic, which lead to an uncontrolled dynamics failing in general to reach the target state.
We will show that this difficulty can be completely avoided by choosing the tracking \eqref{trackingformula-pi} corresponding to the fixed point with $\alpha=\pi$.

Besides the restrictions of crossing imposed in the phase space described below, adiabatic passage needs a sufficiently large pulse area, i.e. $\tau\gg1$; we have determined numerically that the fidelity is already high ($P_{\text{num.}}\gtrsim0.99$) for $\tau\gtrsim 5$.


\subsection{Structure of the instantaneous phase portraits: fixed points and separatrices}

We choose one instantaneous fixed point $P_{\text{track}}(t)$ to construct the detuning $\tD_{\text{track}}(t)$ by the adiabatic tracking formulas \eqref{trackingformula} or  \eqref{trackingformula-pi}. The dynamics defined by $\Omega(t)$ and $\Delta_{\text{track}}(t)$ have other fixed points, that we have to determine in order to analyze their possible effect on the adiabatic evolution. In particular, when some other fixed points are hyperbolic, crossings with  them can be the main obstacle for the adiabatic following of the chosen  $P_{\text{track}}$.
Some general properties of the instantaneous phase portraits of a class of models including the present one were discussed by Itin and Watanabe in \cite{Itin-Watanabe-PRE2007}.  We will present below the particular analysis required for the purpose of adiabatic tracking.

We remark that the continuity of the flow in the reduced phase space  imposes some restrictions on possible crossings of fixed points: An elliptic fixed point can cross a hyperbolic fixed point, but not an elliptic one  (unless it also crosses simultaneously  a hyperbolic fixed point). Furthermore, an elliptic fixed point can cross a hyperbolic one, but it cannot cross any other part of the associated separatrix. In general, when an elliptic and a hyperbolic fixed point cross, they exchange their stability character, the elliptic one becomes hyperbolic and vice-versa.


 \subsubsection{Fixed points}
  $P=1$ is always a fixed point, for any values of the parameters. This fixed point can be stable (elliptic) or unstable (hyperbolic) at different times of the process.
 Once $\tD_{\text{track}}$ is chosen, the  fixed points other than $P=1$ are solutions of the following equations, determined from Eqs. \eqref{Delta-last2}, \eqref{Delta-last2pi}:
 \begin{equation}
 \label{fixed-points-0}
\tD_{\text{track}}\sqrt{P_0}+\tLs \left(\sqrt{P_0}\right)^3+\frac{\tOmega}{2}\left[1-3\left(\sqrt{P_0}\right)^2\right]=0
\end{equation}
 \begin{equation}
 \label{fixed-points-pi}
\tD_{\text{track}} \sqrt{P_\pi} + \tLs \left( \sqrt{P_\pi} \right)^3-\frac{ \tOmega }{2}\left[1-3\left(\sqrt{P_\pi}\right)^2\right]=0.
\end{equation}
These equations are polynomial equations of degree 3 in the variable $\sqrt{P}$. Of the total of six roots, $\sqrt{P_{\text{track}}}$ is one of them, and among the others one has to select the ones that are real and  in the interval $[0,1]$. One can show that the number of such solutions can be $0$, $1$ or  at most $2$. The number of such solutions can vary as a function of time.
\\
At the beginning and at the end of the pulse, when $\tOmega=0$, $P=0$ is a fixed point. Since $P_{\text{track}}(t_i)=0$ and
$P_{\text{track}}(t_f)= 1$, the tracking is the adiabatic following of the family of fixed points that starts as $P=0$.
The tracking fixed point is at all times either at $\alpha=0$ or at $\alpha=\pi$. The fixed points having the other value of $\alpha$ will thus never intersect it.

\begin{figure*}[t]
\begin{center}
\includegraphics[scale=0.95]{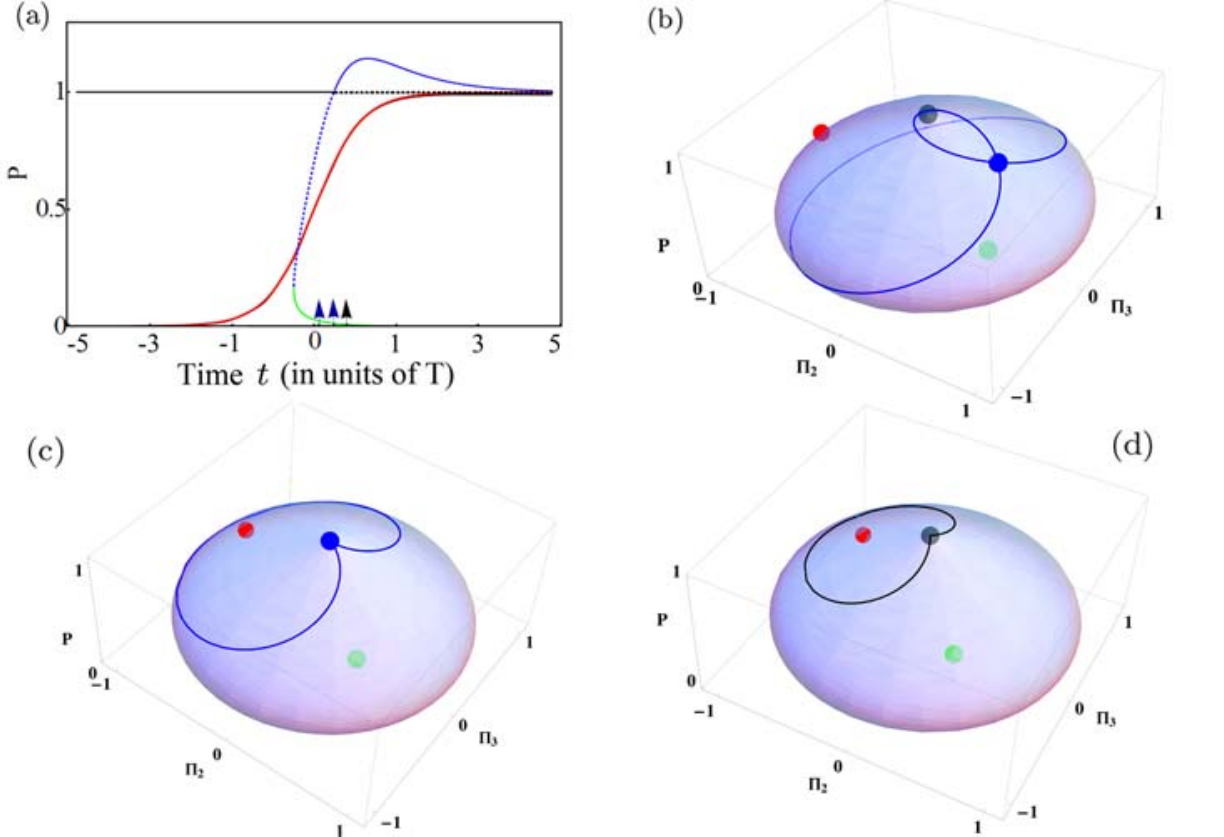}
\end{center}
\caption{Same as Figs. 1 and 2, but for  $\tD_{\text{track}, \pi}$, $\Omega_0/\Lambda_s=0.2$, and $\tau=6$. The exact solution is almost indistinguishable from  $P_{\text{track}}$ (red line). The green and the blue curves are on the same side of the phase space and do not cross the red one, which  is on the opposite side. The (red) tracking fixed point stays elliptic at all finite times.}
\label{fig:AlfaPi_0_2_Fixed_Points}
\begin{center}
\includegraphics[scale=0.95]{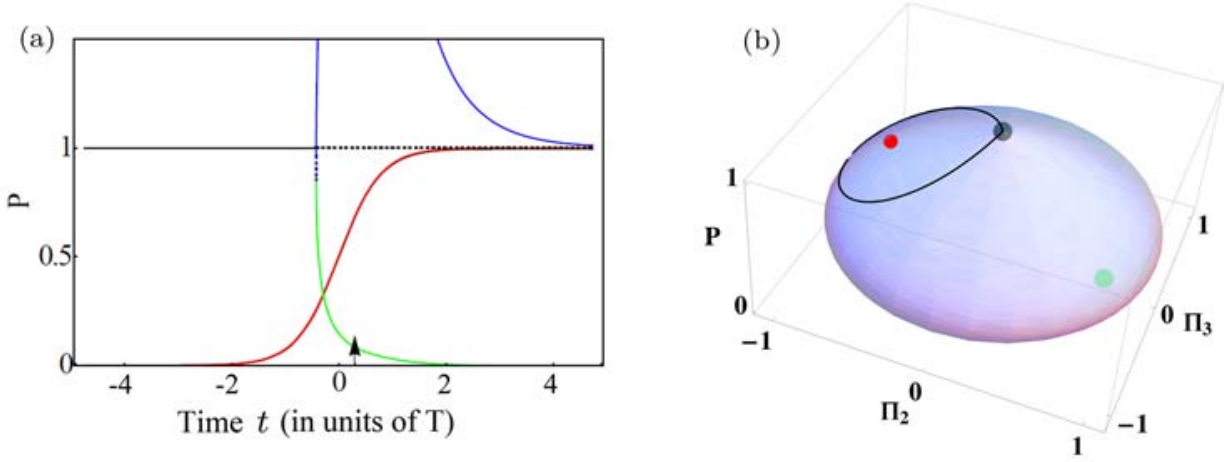}
\end{center}
\caption{Same as Fig. 3 but for $\Omega_0/\Lambda_s=1.1$ and $\tau=9$.}
\label{fig:AlfaPi_1_1_Fixed_Points}
\end{figure*}

\subsubsection{Separatrices}
A separatrix is the energy curve that contains a hyperbolic fixed point.
The energy \eqref{eq: h2new} of a hyperbolic fixed point  $(P_H,\alpha_H)$, in the dimensionless variables \eqref{dimensionless}, is given by (dropping the constant $C$),
\begin{equation}\label{eq: h2new22}
\tilde h_H=\frac{1}{2}\tD P_H+\frac{1}{4}\tLs P_H^{2}+\frac{\tOmega}{2}(1-P_H)\sqrt{P_H}\cos{\alpha_H},
\end{equation}
and thus the points of the  separatrix $(P_S,\alpha_S)$ associated to this hyperbolic fixed point are determined by the relation
\begin{equation}
\tilde h_H= \frac{1}{2}\tD P_S+\frac{1}{4}\tLs P_S^{2}+\frac{\tOmega}{2}(1-P_S)\sqrt{P_S}\cos{\alpha_S}.
\end{equation}
The position of the fixed points and of the separatrices associated with the hyperbolic one organize the global structure of the instantaneous phase portrait in the reduced phase space.

\subsubsection{Evolution of the fixed points and the separatrices for the tracking with
$\tD_{\text{track}, 0}$}
In Figs. \ref{fig:Alfa0_0_5_Fixed_Points} and \ref{fig:Alfa0_1_1_Fixed_Points} we display the time evolution of the instantaneous fixed points for two representative values of  the parameter $\Omega_0/\Lambda_s$. Figure \ref{fig:Alfa0_0_5_Fixed_Points} corresponds to a smaller value of
$\Omega_0/\Lambda_s$ than the one of  Fig. \ref{fig:Alfa0_1_1_Fixed_Points}.

The black full line is the exact  solution $P(t)$ obtained by numerical solution of the differential equations \eqref{eq: ic1-2}.
The blue curve corresponds to the fixed point that has the same value of $\alpha=0$ as the tracking fixed point, drawn in red, while the green curve corresponds to the other value, $\alpha=\pi$. We observe that  for any choice of the parameter $\Omega_0/\Lambda_s$ there is a crossing of $P_{\text{track}}$ with the other fixed point with $\alpha=0$. (The fixed point with $\alpha=\pi$ is on the other side of the phase space, and thus it does not cross the tracking fixed point). Elliptic fixed points are displayed as continuous lines while hyperbolic  ones are displayed with dashed lines.
The system goes through  of a saddle-center bifurcation \cite{Diminnie-Haberman}.
Before the crossing $P_{\text{track}}$ is elliptic, but after the crossing it becomes hyperbolic and it does not satisfy the conditions for the adiabatic theorem anymore \cite{Arnold-Kozlov-Neihstadt-book,Fasano-Marmi-book}. The end of the process is thus always non-adiabatic, which may lead to a deterioration of the transfer to the all-molecules state.

We observed in numerical simulations that if
$\Omega_0/\Lambda_s$ is large enough the crossing happens when the population is close to $P=1$ and in the end there is still a good transfer despite of the loss of adiabaticity.
If $\Omega_0/\Lambda_s \gg 1$ the crossing can be pushed toward the end of the transfer, where
the equation of motion \eqref{eq: I2new}
shows that, because of the factor $(1-P)$, the population $P(t)$ stays constant when  $(1-P)\to 0$, and thus the crossing  has a negligible effect.
This observation is consistent with the expected property that when $\Lambda_s/\Omega$ is small enough one should observe a behavior close to the one of model without Kerr terms.
This is illustrated in  Fig. \ref{fig:Alfa0_1_1_Fixed_Points}.

\subsubsection{Evolution of the fixed points  and the separatrices for the the tracking with $\tD_{\text{track}, \pi}$}
In Figs. \ref{fig:AlfaPi_0_2_Fixed_Points} and \ref{fig:AlfaPi_1_1_Fixed_Points} we display the time evolution of the instantaneous fixed points  for two representative values of  the parameter $\Omega_0/\Lambda_s$.

In the early times of the process the tracking fixed point (shown in red) is the only fixed point (i.e. the only real root of the third order polynomial equations   \eqref{Delta-last2} \eqref{Delta-last2pi}). At a critical time there is a bifurcation in which the two other roots become real, one corresponding to an elliptic fixed point (shown as a green full line)  and the other to a hyperbolic one (shown as an blue dashed line).
The blue curve corresponds to the fixed points that have the same value of $\alpha=\pi$ as the tracking fixed point. The green curve corresponds to the other value $\alpha=0$, i.e. to the fixed point that is on the other side of the reduced phase space and thus does not cross the tracking  fixed point. The full lines represent elliptic fixed points and the dashed lines hyperbolic ones. We observe that the tracking fixed point has no crossings at finite times, and it keeps its elliptic stable nature. Thus the adiabatic theorem applies until the end of the process and we have an efficient adiabatic transfer even for relatively small values of $\Omega_0/\Lambda_s$. This is illustrated in Fig. \ref{fig:AlfaPi_0_2_Fixed_Points}.
The blue full line is the exact  solution $P(t)$ obtained by numerical solution of the differential equations \eqref{eq: ic1-2}.
Figure \ref{fig:AlfaPi_1_1_Fixed_Points} corresponds to a larger value of $\Omega_0/\Lambda_s$.


\section{Conclusion and discussion}
In this work we have analyzed the method of adiabatic tracking for non-linear two-states models of photo- and magneto-association of Bose-Einstein atomic condensates, which include Kerr type nonlinearities \eqref{model}.

We can summarize the main results as follows:
We have first found, via Eq. \eqref{eq: p}, that one cannot achieve a complete transfer with pulses of a finite area, one can only approach it asymptotically, as it was already established for the models without Kerr terms.
The presented analysis shows that a good adiabatic transfer can be achieved by the adiabatic tracking approach, also in the presence of Kerr terms.
It is indeed still possible to construct a detuning that leads to an efficient tracking of the chosen population dynamics despite the strong modifications of the structure of the adiabatic phase portrait, through interfering fixed point trajectories, which induces bifurcation and loss of stability of the adiabatic dynamics.
To show this result, we have first reduced the model \eqref{eq: ib1-2} in order to highlight two relevant combinations of Kerr terms $\Lambda_s$ and $\Lambda_a$. They produce two distinct effects: The term $\Lambda_a$ is a shift that can be easily compensated via a static detuning term. The $\Lambda_s$ term leads to a behavior qualitatively different from the non-linear model without Kerr terms.
We have analyzed the method of adiabatic tracking to design the detuning and to determine the range of pulse peaks, for a given pulse shape, allowing one to approach the complete transfer. We have found two qualitatively different results \eqref{trackingformula} and \eqref{trackingformula-pi} for the detuning.
Only the one with $\alpha=\pi$, \eqref{trackingformula-pi}, leads to a stable adiabatic transfer to the target state, as in the case without Kerr terms. For the other one, with $\alpha=0$, there is an unavoidable crossing of the tracking fixed point with another fixed point, that gives a hyperbolic character to the tracking fixed point, thus destroying the adiabaticity, which can strongly degrade the quality of the transfer for small amplitudes of the driving pulse.
If one takes strong enough pulses, compared with the magnitude of the Kerr terms, the effect of this nonadiabatic crossing can be made negligible. The key result of our paper is that one can always design a path, i.e. a detuning for a given field, that leads to a very efficient association even for strong Kerr terms that can be as large as the corresponding Rabi frequency.

From a practical point of view, in photo- and magneto-association of atomic BEC into molecular BEC, the Kerr terms are proportional to the density while the coupling scales as the square root of the density \cite{Drummond2}.
Some typical values, e.g. for a $^{87}$Rb condensate, are \cite{Mackie-2005}:
\begin{align}
 &\Lambda_{11} =4.96\times10^{-11} \rho~{\rm s}^{-1},  \ \Lambda_{22} = 2.48\times10^{-11} \rho~  {\rm s}^{-1} \nonumber\\
 & \Lambda_{12} =\Lambda_{21} = -6.44\times10^{-11} \rho~{\rm s}^{-1}
 \end{align}
with $\rho$ the density (in cm$^{-3}$), typically $\rho=\rho_0\equiv4.2 \times 10^{14}{ \rm cm}^{-3}  $,
giving
\begin{equation}
\Lambda_s= 2.40\times10^{-10} \rho~{\rm s}^{-1}, \quad \Lambda_{a}=1.64 \times10^{-10} \rho~{\rm s}^{-1}.
\end{equation}
For this example, we have determined numerically that
\begin{itemize}
\item Kerr terms of the order $ \Lambda_s\sim \Omega_0/2$ already lead to an infidelity of 10 $\%$ for the transfer if one does not take into account the compensating term proportional to $\tilde\Lambda_s$ in the design of the detuning \eqref{trackingformula} or \eqref{trackingformula-pi};
\item the distinction between the dynamics induced by the two respective detunings \eqref{trackingformula} and \eqref{trackingformula-pi} can be observed for Kerr terms of the order $ \Lambda_s\sim \Omega_0$;
\item one can compensate the Kerr terms with a high-fidelity transfer using the design \eqref{Delta-last2pi} for Kerr terms of the order $ \Lambda_s\sim 2 \Omega_0$, i.e., for given Kerr terms, the choice of the detuning  \eqref{Delta-last2pi} giving a high fidelity transfer allows a lower peak Rabi frequency, as low as $\Omega_0\sim \Lambda_s/2$.
\end{itemize}
This opens thus the possibility to explore and achieve the association at (i) larger densities than usually used (but not too large so that S-wave scattering is dominant) and/or (ii) lower field amplitudes, by designing the control detuning according to  \eqref{trackingformula-pi}.  Note that, in any case, the field duration $T$ has to be adapted such that $\tau\equiv \Omega_0T \gtrsim 5$ in order to maintain adiabaticity.
For instance, typical coupling for photo-association \cite{Mackie-2005} $\Omega_0=2.1\times 10^6 \sqrt{\rho/\rho_0}~{\rm s}^{-1}$, i.e. $\Omega_0/ \Lambda_s\approx 10^{16} \sqrt{1/(\rho\rho_0)}$, allows one in principle to multiply the density of the condensate by up to 2500 or to divide the field amplitude by up to 50 (or any combination of these) to still reach a high fidelity transfer.

The non-linear two-state system presents limitations to fully describe photo- and magneto-association of degenerate gases. A more realistic model should take into account the fact that the moving non-associated atoms should be described in the continuum of the potential energy of the corresponding molecule (dressed by the trap), which also features rovibrational states \cite{Koch}. The influence of such continuum or/and other possible nearby states should be analyzed in the context of degenerate gases. Furthermore, weakly bound molecule are produced by Feschbach resonance or by direct photoassociation. The resulting molecules are next transferred to ground states by means of Raman processes \cite{Kohler,CChin}. A direct $\Lambda$-photo-association by stimulated Raman adiabatic passage (STIRAP) would be preferred  \cite{Drummond2,Ling}.
Such STIRAP process has the potential to strongly minimize loss and decoherence \cite{Drummond2,Ling}. It has been shown \cite{Mackie-2005} that a low density of a $^{87}$Rb condensate would in principle enhance the molecular conversion efficiency by reducing the Kerr terms compared to a standard density. However, it has been argued \cite{Drummond3} that the reduction of the density causes in general several practical problems. Our alternative strategy to maintain or even increase the density appears thus in principle relevant in this configuration. In future works we will explore the extension of our results in $\Lambda$ systems, taking into account the additional issue that the classical Hamiltonian for the three-state problem is non-integrable. We will also consider the influence of continuum and possible other nearby states in the context of non-linear STIRAP.


The language of photoassociation has been adopted in this paper, however, the derived results are general for nonlinear problems that are described by \eqref{model}, arising also in other physical domains, for instance in nonlinear optics \cite{optics}, such as frequency conversion beyond the undepleted pump approximation \cite{FreqConv}.

The adiabatic tracking ensures a certain robustness of the dynamics and a very good fidelity of the process. Achieving an ultra high fidelity comparable to the one obtained by optimized adiabatic passage of linear systems (see for instance \cite{parallel}) or for shortcut to adiabaticity techniques \cite{shortcut} is an open question.
Studies relevant to these issues have very recently appeared \cite{Dou,XiChen}.

\vfill

\section*{Acknowledgments}

This research has been conducted in the frame of the International Associated Laboratory (CNRS-France and SCSArmenia)IRMAS. We acknowledge additional support from
the European Union Seventh Framework Programme through
the International Cooperation ERAWIDE GA-INCO-295025-
IPERA. C.L. acknowledges the project "Leading Russian Research
Universities" (Grant No. FTI 24 2016 of the Tomsk
Polytechnic University). A.I. acknowledges the support from
the Armenian State Committee of Science (SCS Grant
No. 15T-1C323) and the project "Leading Russian Research
Universities" (Grant No. FTI 24 2016 of the Tomsk Polytechnic
University).M.G. thanks the Cooperation and Cultural Action
Department (SCAC) of the French Embassy in Armenia
for a doctoral grant, as well as the support from the Armenian
State Committee of Science (SCS Grant 14A-1c93 – 2014). The
research undertaken in this publication is supported by a grant of the National Foundation for Science and Technology (NFSAT),
Youth scientists' support program Grant YSSP-13-43.

\end{document}